# Hybrid Quantum Algorithms for Computational Chemistry: Application to the Pyridine–Li$^+$ Complex

Fatemeh Ghasemi,[1] Yousung Kang,[1] Yukio Kawashima,[2] and Kyungsun Moon[1,3]

[1]*Department of Physics, Yonsei University, 03722, Seoul, South Korea*

[2]*IBM Quantum, Tokyo, Japan*

[3]*Institute of Quantum Information Technology,*
*Yonsei University, 03722, Seoul, South Korea*

## Abstract

Accurately capturing electron correlation in large-scale molecular systems remains one of the foremost challenges in quantum chemistry and a primary driver for the development of quantum algorithms. Classical configuration–interaction methods, while rigorous, suffer from exponential scaling, rendering them impractical for large or strongly correlated systems. Overcoming this limitation is central to realizing the promise of quantum computing in chemistry. Here, we investigate the pyridine–Li$^+$ complex using three quantum algorithms: the variational quantum eigensolver (VQE), the subspace quantum diagonalization (SQD) method, and the recently introduced handover iterative VQE (HI-VQE). Our results demonstrate how new generations of hybrid quantum–classical frameworks overcome the scalability and noise sensitivity that constrain conventional VQE approaches. SQD and HI-VQE achieve ground-state energy calculations for problem sizes inaccessible to classical computation, marking a clear advance toward quantum advantage. In particular, HI-VQE enables calculations within active spaces as large as (24e,22o), requiring 44 qubits—well beyond the reach of classical CASCI and VQE. This capability provides a systematic pathway for incorporating increasing numbers of electrons into quantum treatment, thereby approaching exact molecular energies. Importantly, both SQD and HI-VQE exhibit robustness against hardware noise, a critical improvement over earlier approaches. By enabling quantum simulations of molecular systems previously deemed intractable, SQD and HI-VQE offer a realistic route toward practical quantum advantage in computational chemistry. The comparison between HI-VQE and SQD shows that optimizing circuit parameters is crucial for accurate simulation.

# 1  Introduction

Accurate calculation of strongly interacting electron systems remains one of the central challenges in theoretical and computational chemistry. Progress in this area is critical across disciplines, where reliable energy profiles of large molecules and their complexes underpin advances in catalysis, materials science, and drug discovery. Existing computational methods still lack the power to determine the electronic structure of large systems directly from first principles, as solving the fundamental quantum-mechanical equations at scale remains intractable. Consequently, researchers rely on approximate approaches such as Hartree–Fock (HF), density functional theory (DFT), or multiconfigurational schemes like CASCI, each of which captures only part of the electron–electron correlation. To overcome these limitations and extend quantum chemical techniques to larger and more complex systems, embedding strategies have emerged. In these frameworks, the system is partitioned into an active region—treated with a high-level method—and an environment, described at a more approximate level [1, 2].

Quantum computing offers a fundamentally new route for overcoming the limitations of classical simulations of large molecular systems. By encoding quantum states directly in qubits, quantum algorithms avoid the need for many of the approximations inherent in conventional methods [3–6]. Current noisy intermediate-scale quantum (NISQ) devices remain restricted in qubit number and circuit depth, and cannot yet eliminate the exponential scaling problem. Nevertheless, future fault-tolerant quantum computers are expected to enable simulations of much larger molecules than classical approaches can achieve [7–9].

Embedding strategies provide a practical pathway for combining NISQ devices with classical computation. In quantum–classical embedding, strongly correlated subsystems are treated with quantum algorithms, while the remainder of the system is described at a lower classical level [2, 10]. Among these, the variational quantum eigensolver (VQE) has emerged as a leading candidate for near-term applications. However, applying VQE to large active spaces presents a central challenge: accuracy demands a sufficiently large orbital set, but expanding the active space rapidly increases qubit requirements and circuit depth, exceeding the capacity of current hardware [11–13]. Although reliable results have been reported for small molecules [14–19], scaling to larger systems introduces high measurement costs, slow convergence, and strong sensitivity to noise [20].

Accurate determination of molecular energies and potential energy surfaces (PES) is essential for predicting bond strengths, reaction barriers, and equilibrium geometries. To address the limitations of VQE, new hybrid quantum–classical algorithms have been proposed [21, 22]. Notable examples include the Sample-based Quantum Diagonalization (SQD) [23] and the Handover Iterative VQE (HI-VQE) [24]. Unlike standard VQE, which relies on a single variational trial state, these methods construct a small effective Hamiltonian in a carefully chosen subspace of important configurations. Classical diagonalization within this subspace captures electron correlation and excited states more efficiently, reducing measurement overhead and improving robustness to noise. By shifting part of the computational cost from quantum optimization to classical processing, SQD and HI-VQE achieve chemically meaningful accuracy with fewer quantum resources.

In this study, we employ SQD and HI-VQE to investigate the adsorption of $Li^+$ on pyridine. Pyridine ($C_5H_5N$) is a nitrogen-containing heterocycle that binds cations through the lone pair on its nitrogen atom, forming complexes relevant to hydrogen storage catalysis, organic synthesis [25–28] and battery technologies [29–33]. The pyridine–$Li^+$ complex contains 44 paired electrons, which requires 70 spin orbitals on a 6-31G basis [34]—equivalent to 70 qubits. Direct simulation of such a system is infeasible on classical simulators and impractical on current quantum hardware, even with 127 physical qubits, due to noise and limited coherence. We therefore adopt an active-space embedding strategy, treating the essential orbitals with quantum methods while approximating the remainder classically [11, 35].

Our results show that SQD and HI-VQE reproduce classical CASCI energies for large active spaces and remain stable under hardware noise, outperforming conventional VQE. Both methods yield a smooth, chemically meaningful PES with high quantitative accuracy. In particular, HI-VQE enables ground-state energy calculations for active spaces as large as (24e,22o), where classical CASCI and SQD were computationally prohibitive. Ref. [36] shows that SQD can be extended to larger active spaces with tuning of the computational parameters. These advances demonstrate that hybrid quantum–classical subspace methods, combined with embedding, provide a scalable and noise-resilient route toward achieving practical quantum advantage in computational chemistry.

## 2 Theoretical background

Quantum algorithms increasingly draw on techniques from quantum chemistry to represent many-body wavefunctions in quantum circuits. Among these, the Hartree–Fock (HF) method [37] serves as a crucial reference point, providing a mean-field approximation from which more correlated quantum states can be systematically constructed. The HF framework establishes a foundational state for embedding electronic structure problems into the qubit space, leveraging the formalism of second quantization. In practice, the molecular Hamiltonian is first expressed in terms of fermionic creation and annihilation operators, and subsequently mapped onto a qubit Hamiltonian via established transformation schemes [38–41]. This procedure enables an efficient initialization of quantum states and sets the stage for variational and post-HF approaches on quantum hardware.

### 2.1 Constructing the effective Hamiltonian

We will start with the following Hamiltonian to describe a many-body system, which consists of $M$ nuclei and $N$ electrons in atomic units ($\hbar = 1$, $m_e = 1$, $e = 1$) as

$$H = -\sum_i \frac{\nabla^2_{\mathbf{R}_i}}{2M_i} - \sum_i \frac{\nabla^2_{\mathbf{r}_i}}{2} - \sum_{i,j} \frac{Z_i}{|\mathbf{R}_i - \mathbf{r}_j|} + \sum_{i,j>i} \frac{Z_i Z_j}{|\mathbf{R}_i - \mathbf{R}_j|} + \sum_{i,j>i} \frac{1}{|\mathbf{r}_i - \mathbf{r}_j|} \quad (1)$$

where $\mathbf{R}_i$, $M_i$ and $Z_i$ denote the position, mass and charge of the nuclei, respectively, and $\mathbf{r}_i$ is the position of the electrons. This formulation of the Hamiltonian, which ensures the fermionic nature of electrons through an antisymmetric wave function, is commonly referred to as the first-quantized form of the molecular Hamiltonian. For treating our problem on a quantum computer we focus on the second-quantized formulation of the electronic part of the Hamiltonian:

$$H = -\sum_{pq} h_{pq} c^\dagger_p c_q + \frac{1}{2} \sum_{pqrs} g_{pqrs} c^\dagger_p c^\dagger_q c_r c_s \quad (2)$$

where $h_{pq}$ and $g_{pqrs}$ are the integrals of one electron and two electrons, respectively, and $c^\dagger_p$ and $c_p$ are the fermionic operators of creation and annihilation. In an embedding scheme, the molecular orbitals (MOs) are partitioned into active orbitals (AO) and inactive orbitals (IO), which are treated explicitly in quantum simulations and classically, respectively. This approach reduces the number of qubits needed to study the electronic energy of our molecular system. This significantly improves the efficiency of the calculation while maintaining a high degree of accuracy. The geometry of the molecules is shown in Fig. (1). The partition yields two distinct sets of indices, assigned either to $A = \{p|p \in \text{AO}\}$ or to $I = \{p|p \in \text{IO}\}$. In order to construct the Hartree-Fock Hamiltonian for the IO, we need to perform a HF calculation on the IO to get the mean-field approximation. The Fock operator $F_I$ for the inactive orbitals is given as follows:

$$F_I = \sum_{p,q \in I} h_{pq} c_p^\dagger c_q + \sum_{p,q,r \in I} (2g_{prqr} - g_{pqrr}) c_p^\dagger c_q \qquad (3)$$

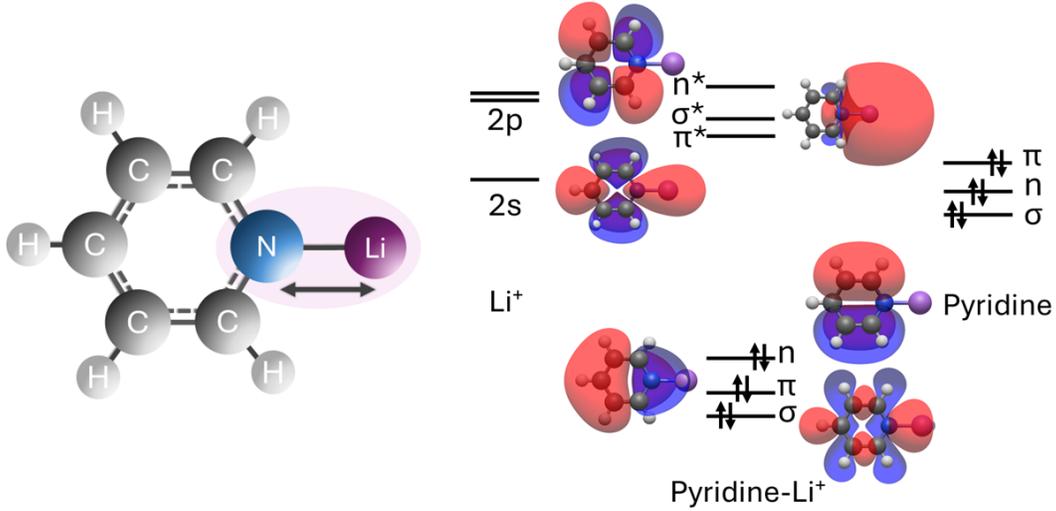

FIG. 1: The geometry (left side) and the molecular orbital diagram (right side) of the pyridine–Li$^+$ complex. Six active orbitals (HOMO, LUMO, HOMO-1, LUMO+1, HOMO-2, LUMO+2) are depicted based on their energies.

This term represents the effective one-electron Hamiltonian for the inactive space, capturing both the direct Coulomb (two-electron repulsion) and exchange interactions. It can be used to construct the effective potential $\tilde{V}_{pq} = \sum_{r \in I}(2g_{prqr} - g_{prrq})$ for $p, q \in A$ that modifies the one-electron part of the Hamiltonian for the active space. The result is that the effective Hamiltonian for the active space includes contributions from both the active orbitals themselves and the mean-field potential from the inactive orbitals, which comes from the Fock

operator:

$$H_{\text{eff}} = \sum_{p,q \in A} (h_{pq} + \tilde{V}_{pq})c_p^\dagger c_q + \sum_{p,q,r,s \in A} g_{pqrs} c_p^\dagger c_q c_r c_s \qquad (4)$$

It should be emphasized that only the inactive orbitals are treated at the mean-field level (Hartree-Fock). The effective resulting Hamiltonian in the active space, $H_{\text{eff}}$, still contains explicit two-electron interaction terms and thus goes beyond a simple mean-field description. We apply the Jordan-Wigner (JW) transformation to map the second-quantized form of the effective Hamiltonian into the qubit space [38, 39].

## 3 Result and discussion

### 3.1 PES for pyridine–Li$^+$ using VQE algorithm

In this section, we report the adsorption of Li$^+$ on the pyridine molecule using an HF-based embedding scheme. We simplify the computational problem by excluding deeply bound core electrons and their associated orbitals from the active space. To further reduce the number of active electrons and orbitals, we define three active spaces: (2e,2o), comprising two electrons in two spatial orbitals; (4e,4o), with four electrons in four spatial orbitals; and (6e,6o), containing six electrons in six spatial orbitals, as illustrated in Fig. 1. This strategy enhances the efficiency of quantum simulations by concentrating computational resources on the most relevant electronic degrees of freedom. Fig. 4 presents the potential energy curves (left panels) and energy differences (right panels) for Li$^+$ adsorption on pyridine, obtained with the 6-31G basis. Results for the active spaces (2e,2o), (4e,4o), and (6e,6o) appear in the top, middle, and bottom panels, respectively. To benchmark the quantum algorithm, we first computed exact values using the classical Full Configuration Interaction (FCI) method. Noiseless statevector-based VQE simulations with the EfficientSU2 ansatz (red markers) converge to the FCI results, confirming the accuracy of the algorithm for this system.

The light purple line in Fig. 4 shows raw results from the ibm_yonsei device without error mitigation. To evaluate the reliability of our approach on current NISQ hardware, we performed ground-state energy computations across all active spaces. For each case, we measured average energies over 10 experiments, both without mitigation (RAW) and with two combined error mitigation schemes (CEM1, CEM2). We compared these results against the FCI reference (black curve). Error bars indicate standard deviations across the 10 runs. Each hardware experiment employed 10,000 shots per measurement.

Because unmitigated quantum computations deviated significantly from the FCI reference, error mitigation proved essential. We systematically explored techniques individually and in combination. The lowest errors arose when dynamical decoupling (DD), Pauli twirling, TREX, and zero-noise extrapolation (ZNE) were applied together [47]. Pauli twirling converts coherent noise into Pauli noise, reducing its accumulation rate. TREX corrects readout errors, improving measurement accuracy. ZNE further mitigates errors through extrapolation, using quadratic fits over noise factors 1, 2, and 4, and linear fits over factors 1, 2, and 3. To suppress qubit decoherence from environmental noise, we applied DD with the XY4 sequence.

We implemented two combined error mitigation schemes, CEM1 and CEM2, which both included DD, Pauli twirling, and TREX, but differed in the ZNE method. CEM1 employed linear extrapolation over noise factors 1, 2, and 3, whereas CEM2 used quadratic extrapolation over factors 1, 2, and 4. To benchmark under consistent noise conditions, we also used the Qiskit "Fake Sherbrooke" backend (blue markers in Fig. 2), enabling direct comparison with real hardware results.

Fig. 2 shows that, across all active spaces, statevector simulations (red markers) agree closely with FCI along the adsorption profile. In contrast, unmitigated quantum device results (RAW) deviate substantially: 60 mHa for the (2e,2o) active space, 350 mHa for (4e,4o), and 800 mHa for (6e,6o).

Both CEM1 and CEM2 markedly reduce these discrepancies. CEM1 yields highly consistent results across 10 independent measurements, with minimal dispersion and small standard deviations, enabling reliable data fitting and parameter estimation. In comparison, CEM2 produces average energies that differ slightly from the reference but exhibit significant variability over measurement. For (4e,4o) and (6e,6o), fluctuations in mean values make it difficult to obtain a reliable estimate of the parameters. Consequently, CEM1 emerges as the more robust and reliable error mitigation strategy for this system.

The growing deviation from FCI with increasing active space reflects both computational complexity and hardware limitations. Larger active spaces demand more qubits and gates, deepening circuits and amplifying noise, gate errors, and decoherence. These effects drive deviations from 60 mHa in (2e,2o) to 800 mHa in (6e,6o), underscoring the current con-straints of NISQ devices in quantum chemistry. Because RAW results diverge strongly from FCI, the vertical scale of the PES is stretched, making the profiles for (4e,4o) and (6e,6o) appear artificially flat.

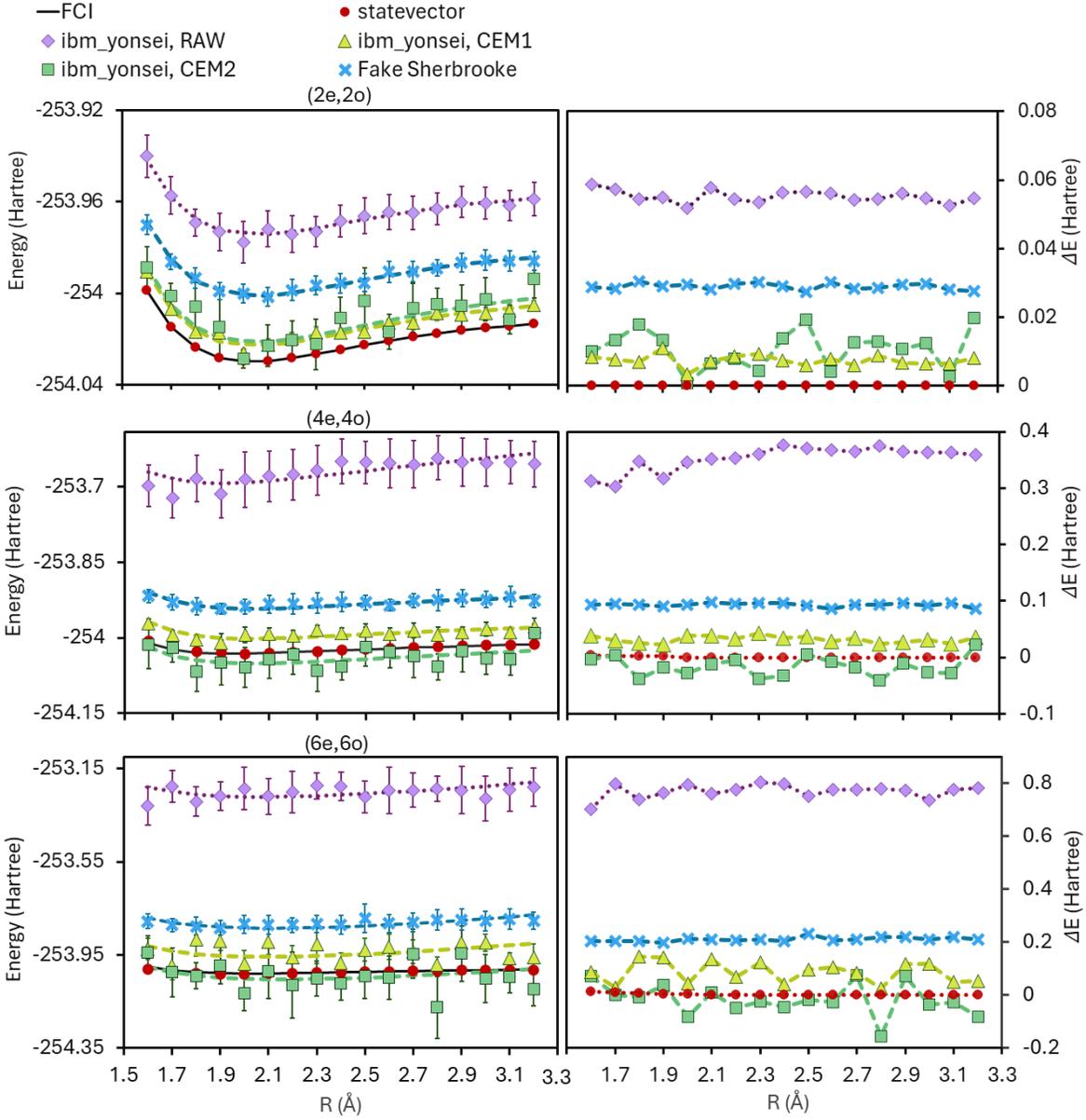

FIG. 2: Potential energy surfaces for the pyridine–Li$^+$ interaction (left Panels), and energy deviations from FCI in Hartree units (right panels), obtained using the embedding approach with various computational methods for the (2e,2o) (top panels), (4e,4o) (middle panels) and (6e,6o) (bottom panels) active spaces using the 6-31G basis set. Error bars show the standard deviation of 10 independent experiment repetitions and the coordination of the markers correspond to the average value obtained from various computational methods.

As shown in the right panels, application of combined error mitigation reduces deviations dramatically: from 60 mHa to 1 mHa for (2e,2o), 350 mHa to 4 mHa for (4e,4o), and 800 mHa to 10 mHa for (6e,6o). These results highlight the critical role of error mitigation in enabling accurate quantum simulations on current quantum computing hardware.

## 3.2 Potential Energy Surface for Higher Active Spaces Using SQD and Hi-VQE Methods

In this section, we present large active space embedding calculations using the quantum SQD and HI-VQE methods as subsystem solvers. Fig. 3 shows the potential energy surface of the pyridine-Li$^+$ complex obtained with the SQD and HI-VQE methods compared to classical RHF and CASCI reference calculations. The active space consists of 16 active spatial orbitals represented by 32 qubits, where 12 active electrons will occupy. The 16 spatial orbitals below the active orbitals are fully occupied by 32 frozen electrons. This corresponds to the (12e,16o) active space. As expected, the RHF curve lies higher in energy due to the lack of dynamic correlation, while CASCI provides a reliable reference within the chosen active space. The right panel shows how much the SQD and HI-VQE energies differ from the CASCI reference. The SQD results closely follow the CASCI trend but show small systematic deviations on the order of $8 \times 10^{-3}$ Ha. This difference originates from the finite sampling and subspace truncation used in the SQD method. The energy difference between HI-VQE and CASCI remains consistently below $5 \times 10^{-4}$ Ha across all bond distances. This deviation is an order of magnitude smaller than the conventional threshold of chemical accuracy ($\sim 1.6 \times 10^{-3}$ Ha), demonstrating that HI-VQE is capable of reproducing classical active-space results with high quantitative precision. The robustness of HI-VQE establishes it as a reliable approach for chemically accurate potential energy surfaces on near-term quantum hardware. Its iterative handover mechanism improves the consistency and accuracy of subspace-based quantum algorithms, enabling PES reconstruction within the limits of NISQ devices. Importantly, the smoothness of the HI-VQE curve confirms that iterative subspace diagonalization correctly incorporates and combines multiple key electronic configurations into the effective Hamiltonian of the many-body molecular system.

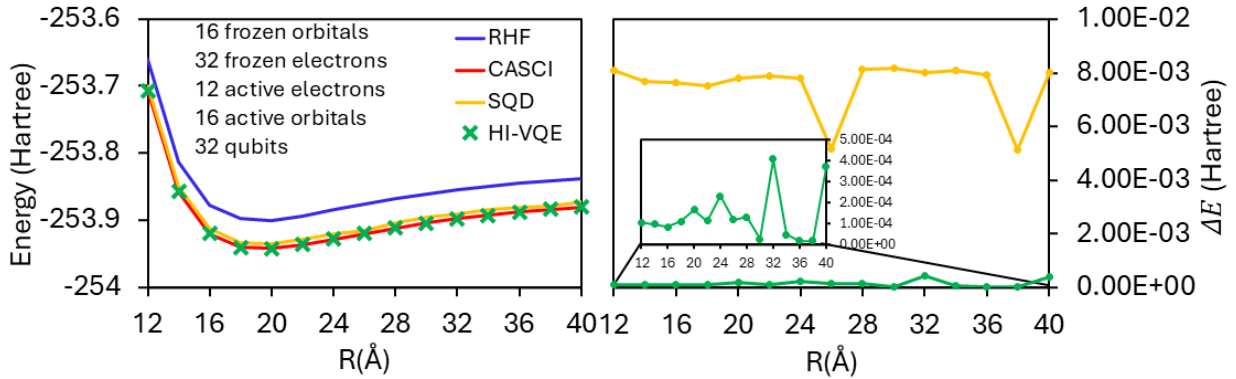

FIG. 3: Potential energy curves (left) and energy deviations (right) for the pyridine-Li$^+$ complex obtained using RHF, CASCI, SQD, and HI-VQE methods. Calculations were performed with 12 active electrons mapped to 32 qubits: (12e,16o). The left panel shows the total electronic energy as a function of the Li–N bond distance ($R$), while the right panel presents the energy deviations $\Delta E = E_{\text{SQD/HI-VQE}} - E_{\text{CASCI}}$. The HI-VQE results closely reproduce the CASCI reference with deviations below $5 \times 10^{-4}$ Ha, whereas the SQD energies show slightly larger errors ($\sim 8 \times 10^{-3}$ Ha). These results confirm that HI-VQE achieves high quantitative agreement with CASCI results, demonstrating its robustness and reliability compared to both SQD and classical reference methods.

Using the HI-VQE and SQD methods, we calculated the energies of the pyridine-Li$^+$ complex within an active space of 16 active electrons in 18 active spatial orbitals (16e,18o), mapped to 36 qubits. SQD energies lie slightly above those from HI-VQE, with deviations on the order of $8 \times 10^{-3}$ Ha, reflecting the improved accuracy achieved through iterative refinement in HI-VQE. Crucially, this calculation lies beyond the reach of CASCI, as the exponential scaling of the determinant space renders classical diagonalization infeasible. In the absence of a classical reference, these results mark entry into a regime approaching quantum advantage.

Extended studies with the HI-VQE method demonstrate a significant advance in the scalability of quantum simulations. We successfully evaluated ground-state energies for a system containing 24 active electrons, mapped to 44 qubits—a size at which both CASCI and SQD methods become computationally intractable. Nevertheless, Ref. [36] demonstrates that access to larger active spaces can be achieved in SQD through a careful parameter tuning. This achievement marks a meaningful development in computational chemistry, showing that the hybrid HI-VQE algorithm can access system sizes far beyond the reach of classical approaches.

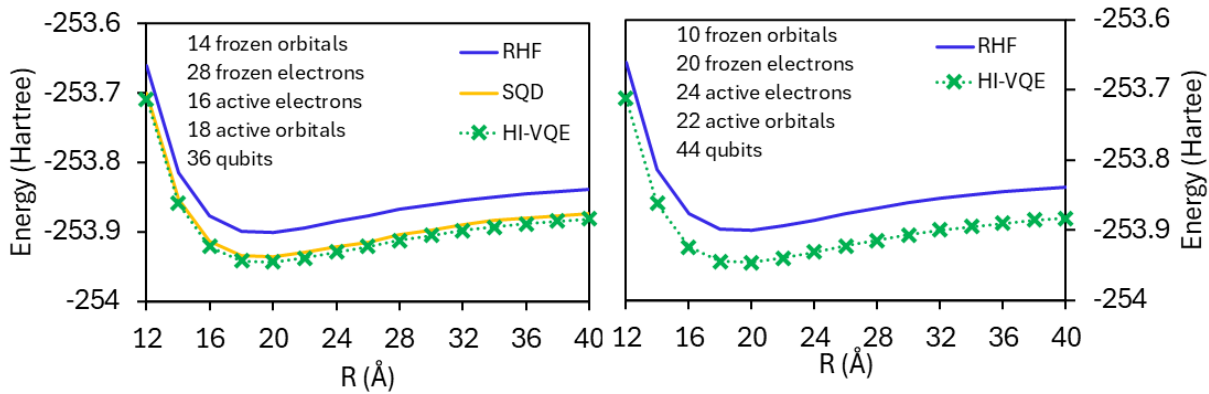

FIG. 4: Ground-state energy profiles of the pyridine-Li$^+$ complex obtained using RHF, SQD, and HI-VQE methods for two different active-space configurations. The left panel corresponds to 16 active electrons in 18 active spatial orbitals (16e,18o), mapped onto 36 qubits, while the right panel shows results for 24 active electrons in 22 active spatial orbitals (24e,22o), mapped onto 44 qubits. With the exponential increase in configuration space size, CASCI calculations become impractical for both configurations, and SQD also fails to produce converged results for the larger active-space system. In contrast, the HI-VQE method successfully produces smooth and stable potential energy curves, demonstrating its scalability and reliability for systems beyond the reach of classical and other sample-based quantum algorithms.

HI-VQE generates robust, smooth, and stable potential energy surfaces for systems as large as (24e,22o), accurately capturing complex electron correlation effects. This capability represents a notable step toward practical quantum advantage in electronic structure calculations. The broader significance of this achievement lies in demonstrating that quantum algorithms can now address realistic molecular systems too large or strongly correlated to be treated exactly by classical simulations such as CASCI, FCI, or other diagonalization-based approaches. In HI-VQE, computational cost is reduced by restricting the Hilbert space, i.e., the number of electronic configurations considered in the quantum calculation. While such approximations may introduce errors, HI-VQE maintains high precision by iteratively refining the selected subspace during the handover process. As system size increases, the method incorporates more configurations while preserving efficiency, enabling accurate calculations

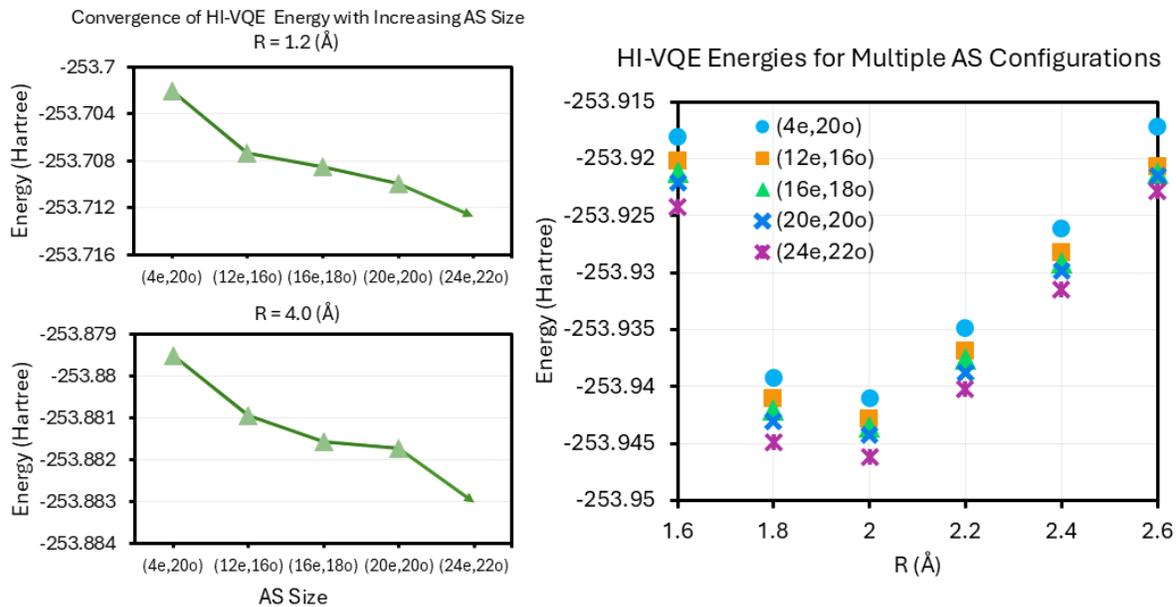

FIG. 5: HI-VQE ground-state energies for the pyridine–Li$^+$ complex using different active-space sizes. The results show that HI-VQE remains stable even when the active space becomes large, allowing a large fraction of the total electrons to be included in the calculation. Including more electrons in the active space allows HI-VQE to capture more electron correlation, which lowers and stabilizes the energy. This result highlights that HI-VQE empowers quantum simulations to approach the exact, full-electron solution by enabling the treatment of large and correlated electronic systems that classical methods fail to treat.

on near-term quantum devices.

A key advantage of HI-VQE is its ability to treat large active spaces, enabling inclusion of most electrons in the pyridine–Li$^+$ complex (Fig. 5). As the active space expands, HI-VQE energies approach those expected from full-electron calculations. In our study, the algorithm successfully handled an active space of (24e,22o), a scale at which classical methods such as CASCI become entirely impractical. Despite the size and complexity of this configuration space, HI-VQE remained stable and quantitatively reliable, recovering a substantial portion of the electron correlation. Ref. [36] demonstrates that orbital and circuit parameter optimization can improve the SQD result and access to larger active spaces can be achieved in SQD through a careful parameter tuning.

This capability provides a realistic pathway for quantum computers to approximate exact molecular energies by systematically incorporating more electrons into quantum treatment.

These results offer one of the first clear indications that hybrid quantum–classical algorithms are beginning to access problem sizes beyond the reach of classical methods, marking an important step toward practical quantum advantage in computational chemistry.

## 4 Conclusion

In this work, we applied advanced hybrid quantum–classical algorithms—including VQE, SQD, and HI-VQE—to the pyridine–$Li^+$ complex characterized by a dative interaction. Although conventional VQE was limited to small active spaces, both SQD and HI-VQE enabled ground-state energy calculations in substantially larger spaces, extending beyond the reach of classical configuration interaction methods. Notably, HI-VQE allowed evaluation of ground-state energies for molecular configurations with active spaces as large as (24e,22o), even in cases where classical CASCI results were unattainable. By selectively sampling the most relevant components of the wavefunction, SQD and HI-VQE circumvent the exponential scaling inherent to classical diagonalization, achieving a favorable balance between efficiency and accuracy. The resulting energies were used to construct the potential energy surface of the pyridine-$Li^+$ complex, demonstrating that HI-VQE can produce chemically meaningful predictions for systems inaccessible to classical approaches.

Beyond scalability, HI-VQE exhibited notable resilience to hardware noise: its iterative handover mechanism mitigates error accumulation, enabling stable and reliable energy estimation on NISQ devices. By providing accurate potential energy surfaces, the method facilitates the prediction of key chemical properties, including equilibrium geometries, binding energies, and reaction characteristics.

This work is consistent with the study in [36] showing circuit parameter optimization is inevitable for accurate simulation. Our results establish SQD and HI-VQE as scalable and robust approaches for probing strongly correlated molecular systems, marking a significant step toward practical quantum advantage in quantum chemistry. SQD and HI-VQE offer a systematic framework that can serve as a benchmark for future quantum-computing studies, providing valuable insight into the current capabilities of quantum algorithms and their potential to address increasingly complex problems in computational chemistry.

# 5 Methods and computational techniques

In this section, we first describe the computational procedure used to construct the effective Hamiltonian. We then outline the quantum-computing components required to obtain ground-state energies with the three algorithms employed, followed by the specific computational settings used for each method.

We used VQE, SQD, and HI-VQE approaches to investigate the pyridine–Li$^+$ complex. The optimized geometry of the pyridine molecule is obtained with DFT calculations at the B3LYP/STO-3G [42] level of theory using the Orca quantum chemistry software [43]. For the further energy calculations we used the 6-31G basis sets [34]. We use PySCF [44] as the classical computing framework, allowing us to integrate both classical and quantum components within the same Python [45] programming environment and accelerate the development process. In all simulations, we use the JW transformation method to map fermionic operators into qubit operators using Pauli matrices. We performed hardware experiments on ibm_yonsei, an Eagle processor with 127 qubits. The Qiskit IBM Runtime service was used via version 1.2.4 of the `qiskit-ibm-runtime` package. The details of each method are outlined in the following subsections.

## 5.1 Circuit design using VQE algorithm

We first combine the VQE algorithm with active space quantum embedding schemes, which defines the subset of important orbitals to be handled by the quantum algorithm.

We construct the ground state trial wave function of a molecule using adjustable parameters within a specifically designed quantum circuit. The Hamiltonian is decomposed into a weighted sum of Pauli operators, and the quantum computer is used only to evaluate expectation values of these terms with respect to a parameterized trial state. The algorithm does not require explicit construction of the Hamiltonian matrix in the active space; instead, a classical optimizer iteratively updates the circuit parameters to find the quantum state that minimizes the energy of a molecular system.

The electronic wave function is presented using EfficientSU2 ansatz [15] as a Hardware-Efficient ansatz. These ansatze consist of repeated, compact blocks of a limited set of parameterized gates that are efficient to implement on current quantum hardware. The ansatz can be written mathematically as follows:

$$U(\vec{\theta}) = \prod_{k=1}^{p}(\prod_{i=1}^{n} R_y(\theta_i^{2k-1}) R_x(\theta_i^{2k})) U_{\text{ent}} \qquad (5)$$

where, $R_y(\theta)$ and $R_x(\theta)$ are the single-qubit rotation gates for each qubit $i$. Moreover, $n$ and $p$ are the number of qubits and repetitions of the ansatz, respectively. In our simulations, we set $p = 1$ that allowed convergence of the VQE energy to $10^{-4}$ Ha precision for the chosen active space. $U_{\text{ent}}$ denotes entangling gates (CNOT) between pairs of qubits. These types of ansatze aim to construct flexible trial states while minimizing the number of gates used, making them ideal for today's quantum devices, which are constrained by short coherence times and limited qubit connectivity. The classical optimizer L-BFGS-B is employed for optimizing the ansatz parameters [46].

The active space embedding approach was implemented in Qiskit ecosystem using the open-source Qiskit Nature framework [47, 48]. We used the exact diagonalization solver (`NumPyMinimumEigensolver` from Qiskit Algorithms) to obtain reference ground-state energies for benchmarking the active space embedding scheme. In addition, noiseless VQE simulations were performed with the statevector backend to test and validate the VQE approach. With a suitable wavefunction ansatz, the VQE is expected to converge to the same results as exact diagonalization. The optimal point obtained from the VQE run using the statevector method was used to prepare the circuit for a single-point energy measurement on real quantum hardware. The quantum circuit was then transpiled with optimization level 3, applying aggressive optimizations to minimize gate depth and enhance performance. To meet the hardware connectivity and topology requirements, we applied the layout from the transpiled circuit to the operator, ensuring that the qubit mapping aligned with the optimized circuit. Additionally, dynamical decoupling was employed to mitigate decoherence effects during circuit execution. Readout error mitigation (ROEM) was also applied to correct measurement errors, both of which are features supported within the Qiskit Runtime environment [49]. To further reduce the impact of hardware noise, we implemented zero-noise extrapolation (ZNE) [50], which involves executing quantum circuits at varying noise levels and extrapolating the results to estimate the ideal, noise-free outcome. Based on the integrated implementation of these techniques, we define two combined error mitigation schemes, CEM1 and CEM2, whose specific configurations are described in Section IV. Resource estimates for each active space are summarized in Table I.

TABLE I: Resource per active space, $p=1$, full entanglement.

| Active Space | Qubits | Params | Depth (ucx) | CNOTs (ucx) | Depth (yonsei) |
|---|---|---|---|---|---|
| (2e,2o) | 4 | 32 | 15 | 9 | 49 |
| (4e,4o) | 8 | 64 | 19 | 21 | 71 |
| (6e,6o) | 12 | 96 | 23 | 33 | 85 |

ucx = decomposed to $\{u, cx\}$; yonsei = post-transpile to *ibm_yonsei*.

## 5.2 Circuit design using SQD algorithm

In the second step, we use the SQD method, which builds an effective Hamiltonian from sampled quantum states and solves it classically to improve the description of electron correlation in the active space. SQD, as a hybrid quantum–classical computational framework, is expected to overcome the exponential complexity of exact diagonalization. In this method, ibm_yonsei hardware is employed to prepare the ansatz state and to generate bitstring samples through repeated projective measurements on the computational basis. Each bitstring corresponds to a configuration that represents a physically meaningful occupation of molecular orbitals. From these measurement outcomes, it is possible to identify and select a subset of configurations that span the reduced subspace of the full Hilbert space. The active-space Hamiltonian is then mapped onto the sampled subspace, and the matrix elements between the basis states, $\langle i|H|j\rangle$, are estimated using quantum measurements. In this way, SQD explicitly constructs a reduced Hamiltonian matrix, and the resulting effective Hamiltonian is subsequently diagonalized on a classical computer. This method uses the strength of quantum computers to sample important states while leaving the computationally expensive diagonalization work to classical computers. Consequently, it can provide approximate values for the ground-state energy and a few excited states of the system.

Here, we calculate ground-state energies as a function of the interatomic distance N – Li in the range of 1.2 to 4 Å. For SQD calculations, we employed a spin-balanced Local Unitary Cluster Jastrow (LUCJ) ansatz to generate bitstring samples. For each geometry point, $10^5$ projective measurement shots were collected, and 50 independent SQD runs were performed by resampling the measured bitstrings to generate a statistical distribution of energies. For each SQD run, we used 5 independent batches, each batch constructed from $10^5$ postselected projective measurement samples. These samples were filtered to enforce the correct numbers of $\alpha$ and $\beta$ electrons and subsequently subsampled to form the individual batches. After removing duplicate configurations, each batch yielded an effective determinant subspace whose dimension is significantly smaller than the number of samples and varies across batches and self-recovery iterations. In the present work, the resulting subspace dimensions typically ranged from $7.6\times 10^4$ to $1.3\times 10^5$ determinants. Configuration-recovery was applied at every iteration after the first, using the average orbital occupations obtained from the previous iteration as input to the carryover procedure. A fixed number of 10 self-consistency iterations was performed for each geometry point. The SQD ground-state energy reported corresponds to the minimum value obtained across the final batch distribution, ensuring that the lowest-energy physically relevant configuration sampled by SQD is captured.

In this work, we tested several combinations of active electrons and orbitals, by increasing the size of the active space to examine how many active electrons from the pyridine-Li$^+$ system could be treated within the SQD framework. The ground-state energy at each distance is extracted from the minimum value of the corresponding distribution, which reflects the most physically relevant low-energy configuration sampled by SQD. This analysis allows us to extend beyond the limitation of the VQE calculations, which were restricted to an active space of (6e,6o). For comparison, the same geometries were computed with Restricted Hartree–Fock (RHF) and Complete Active Space Configuration Interaction (CASCI) methods using the identical active-orbital list, and all three curves (RHF, CASCI, SQD) were plotted in Figs. 3 and 4 to assess the PES.

## 5.3 Circuit design using HI-VQE algorithm

As the final part of our investigation, we use the recently developed HI-VQE algorithm to study the electronic structure of the pyridine–Li$^+$ system. HI-VQE is a hybrid quantum–classical algorithm that is expected to overcome the limitations of conventional variational approaches in strongly correlated electronic systems [24]. In contrast to VQE, which requires repeated Pauli-word measurements and direct optimization of circuit parameters to approximate ground-state, HI-VQE employs an iterative "handover" process between quantum and classical resources. At each iteration, a quantum device prepares a parametrized state and produces samples of electron configurations via projective measurements in the computational basis. These bitstring samples correspond to the Slater determinants $|\psi_i\rangle$ that define a trial subspace $C'_i$ of the full Hilbert space. The effective Hamiltonian matrix elements on this basis are defined as

$$(H_{\text{eff}})_{ij} = \langle\psi_i|\hat{H}|\psi_j\rangle, \tag{6}$$

The classical diagonalization of the projected Hamiltonian yields an approximate ground state $|\Psi\rangle$ of the form

$$|\Psi\rangle = \sum_i \omega_i |\psi_i\rangle, \tag{7}$$

where the amplitudes $\omega_i$ are determined exactly within the subspace $C'_i$. In each iteration, invalid configurations are filtered, low-weight determinants are discarded, and additional excitations are introduced, ensuring that the trial subspace is refined towards the true core space $C$.

The optimization cycle proceeds as follows: (i) prepare a quantum state $|\psi(\boldsymbol{\theta})\rangle = \hat{U}(\boldsymbol{\theta})|\psi_{\text{HF}}\rangle$, (ii) the quantum computer measures this state many times to collect bitstrings $\{b_i\}$ to construct the trial subspace $C'_i$ of important electronic configurations, (iii) project $\hat{H}$ into $C'_i$ and perform a classical diagonalization on a classical computer to obtain an improved energy estimate, and (iv) update $\boldsymbol{\theta}$ using the resulting energy $E_i$ before handing the refined state to the next iteration. The process continues until the energies and subspaces stop changing, which means that the method has converged to a stable and accurate solution that approximates the core space $C$.

Unlike conventional VQE, which requires quantitative accuracy of amplitudes $\omega_i$ from the quantum device at each step, HI-VQE only demands that the important determinants are sampled with sufficient frequency, while their exact amplitudes are reconstructed via classical diagonalization. HI-VQE needs fewer measurements than standard VQE, so it is faster and more practical on real quantum hardware. This method avoids difficult optimization problems and still produces accurate wavefunctions. Because of this, it is especially good for molecules that require many configurations to be described correctly.

Here, HI-VQE algorithm was executed using Qunova's `hivqe-chemistry` function available through the QISKIT FUNCTIONS CATALOG.

We evaluated the pyridine-$Li^+$ potential energy surface by scanning 15 optimized geometries obtained from DFT calculation with the N – Li interatomic distance in the range of 1.2 to 4

Å. For each geometry, molecular integrals were built in PySCF with the 6-31G basis and total charge +1. An active space was specified by freezing inactive orbitals and correlating active orbitals, and this setup was passed to Qunova's HI-VQE chemistry function loaded from the Qiskit Functions Catalog. HI-VQE calculations were performed using a 6-31G basis set and an active-space embedding with frozen core orbitals. The Excitation-Preserving Ansatz (EPA) with circular entanglement and two repetitions was employed, using 100 measurement shots per circuit and up to 30 handover iterations. The effective subspace was truncated to at most $4 \times 10^5$ states (upper bound on the effective Hilbert space dimension), with up to 100 new configurations added per iteration, and all calculations were executed on the ibm_yonsei backend. In the HI-VQE algorithm, a configuration recovery process is incorporated to correct corrupted states using a target electron occupation through carryover mechanism proposed in [23]. For comparison, the same geometries were computed with RHF and CASCI (PySCF) using the identical active-orbital list, and all three curves (RHF, CASCI, HI-VQE) were plotted in Figs. 3, 4 and 5 to assess the PES.

## Acknowledgment

This research was supported by the National Research Foundation of Korea (NRF) through grants NRF-2022M3K2A108385813 and RS-2023-00257561. The authors gratefully acknowledge the funding support that made this work possible.